# Weakness Analysis and Improvement of a Gateway-Oriented Password-Based Authenticated Key Exchange Protocol

He Debiao, Chen Jianhua, Hu Jin

School of Mathematics and Statistics, Wuhan University, Wuhan, Hubei 430072, China

hedebiao@163.com

**Abstract**: Recently, Abdalla et al. proposed a new gateway-oriented password-based authenticated key exchange (GPAKE) protocol among a client, a gateway, and an authentication server, where each client shares a human-memorable password with a trusted server so that they can resort to the server for authentication when want to establish a shared session key with the gateway. In the letter, we show that a malicious client of GPAKE is still able to gain information of password by performing an undetectable on-line password guessing attack and can not provide the implicit key confirmation. At last, we present a countermeasure to against the attack.

**Categories**: D.4.6; C.2.1

## 1. Introduction

The gateway-based authenticated key exchange (GAKE) protocols are important cryptographic techniques for secure communications. Conceptually, a typical three-party password-based authenticated key exchange protocol works as follows. As requirement, each client $C$ shares a human-memorable password with a trusted server so that they can resort to the server $S$ for authentication when want to establish a shared session key with the gateway $G$. Among the various means of authentication that can be considered, the most interesting one from a practical point of view is the password-based setting in which a simple human-memorizable secret, called a password, is used for authentication.

In 2005, Abdalla et al. proposed the first gateway-oriented password-based authenticated key exchange (GPAKE) scheme among a client, a gateway, and an authentication server [1]. Even though Abdalla et al. had proved the session key semantic security of their scheme in a formal model, Byun et al. reported an undetectable on-line guessing attack on the GPAKE protocol where a gateway can iteratively guess a password and verify its guess without being detected by the server [2]. Byun et al. also proposed an improved scheme to eliminate the security vulnerability of Abdalla et al.'s scheme. However, Wu et al. [3] found that Byun et al.'s scheme still cannot resist the on-line undetectable guessing attack.

Very recently, Abdalla et al. [5] present a new variant of the GPAKE scheme of Abdalla et al. [1]. They used the Schnorr's signature [6, 7] in the new scheme in order to guarantee the security of the new scheme. The new scheme can withstand the attack by Byun et al. [2].

In this letter, we review Abdalla et al.'s new protocol [5], and show that it does actually leak information of password to a malicious client and can not provide the implicit key confirmation. Especially, we show that Abdalla et al.'s new scheme is susceptible to an undetectable on-line password guessing attack by a malicious client. We also give a countermeasure against the attack by letting the client generate a message authentication code of keying material.

## 2. Review of Abdalla et al.'s protocol

In this section, we will review Abdallar et al.'s protocol. First we introduce some notations used in our paper. In order illustrate the protocol clearly, some notations are introduced as follows:

- $C$, $G$ and $S$ denote the client, the gateway and the trusted server separately.
- $pw_C$ denotes the password shared between $C$ and $S$.
- $ID_C$ and $ID_G$ denote the identity of $C$ and $G$ separately.
- $\mathbb{G}$ denotes a finite cyclic group having a generator $g$ of bit prime order $q$.
- $sk$ denotes a session key generated between the client and the gateway.
- $h_1(\cdot)$, and $h_2(\cdot)$ denote two secure hash function, such as $SHA-1$.
- $H$ denotes a secure hash function, where $H(\cdot):\{0,1\}^* \rightarrow \mathbb{G}$.
- $NIZKPDL(m; g, h)$ denotes the Schnorr's signature [6, 7] on the message $m$.

In Abdallar et al.'s protocol, each client shares a human-memorable password with a trusted server. When a client wants to establish a shared session key with a gateway, they resort to the trusted server for authenticating each other. Abdallar et al.'s protocol will be described as follows.

**Step 1**: $C$ chooses two random numbers $x$ and $r_C$. Then $C$ computes $X^* = g^x \times H(ID_C, ID_G, pw_C)$, then sends $M_1 = \{ID_C, X^*\}$ to $G$.

**Step 2**: Upon receiving the message $M_1$, $G$ sends $M_2 = \{ID_C, ID_G, X^*\}$ to the server $S$.

**Step 3**: Upon receiving the message $M_2$, the $S$ generates a random number $s$, and computes $\overline{X} = (X^* / H(ID_C, ID_G, pw_C))^s$, $h = g^s$ and $\pi_1 = NIZKPLD(X^*; g, h)$. Then $S$ sends $M_3 = \{\overline{X}, h, \pi_1\}$ to $G$.

**Step 4**: When $G$ receives $M_3$, he/she generates a random number $y$ and computes $\overline{Y} = h^y$, $\pi_2 = NIZKPLD(X^*; g, \overline{Y})$, $K_G = (\overline{X})^y$, $AuthG = h_2(ID_C, ID_G, X^*, \overline{Y}, K_G)$ and the session key $sk_G = h_1(ID_C, ID_G, X^*, \overline{Y}, K_G)$. Then $G$ sends $M_4 = \{ID_G, h, \overline{Y}, AuthG, \pi_1, \pi_2\}$ to $C$.

**Step 5**: After receiving $M_4$, $C$ computes $K_C = (\overline{Y})^x$ and checks weather $AuthG$

equals $h_2(ID_C, ID_G, X^*, \overline{Y}, K_C)$. If not, $C$ stops the session. Otherwise, $C$ checks weather both of $\pi_1, \pi_2$ is valid. If not, $C$ stops the session, else $C$ computes the session key $sk_C = h_1(ID_C, ID_G, X^*, \overline{Y}, K_C)$.

## 3. Security analysis

### 3.1. Undetectable on-line guessing attack

Due to the low entropy, password-based authenticated key exchange protocols suffer from so-called exhaustive dictionary attacks. The attacks on PAKE schemes can be classified into three types [10]:

1) **Off-line dictionary attacks**: an attacker uses a guessed password to verify the correctness of the password in an offline manner. The attacker can freely guess a password and then check if it is correct without limitation in the number of guesses.

2) **Undetectable on-line dictionary attacks**: an attacker tries to verify the password in an on-line manner without being detected. That is, a failed guess is never noticed by the server and the client, and the attacker can legally and undetectably check many times in order to get sufficient information of the password.

3) **Detectable on-line dictionary attacks**: an attacker first guesses a password, and tries to verify the password using responses from a server in an on-line manner. But a failure can be easily detected by counting access failures.

In the following, we demonstrate an undetectable on-line dictionary attack against the Abdalla et al.'s scheme [5] where an adversary is able to legally gain information about the password by repeatedly and indiscernibly asking queries to the authentication server. We assume that $A$ has total control over the communication channel between the user $C$ and the gateway $G$, which means that he/she can insert, delete, or alter any messages in the channel. The detailed description of the attack is as follows:

**Step1**. $A$ guesses a password $pw_C'$ from a uniformly distributed dictionary $D$ and computes $PW_C' = H(ID_C, ID_G, pw_C')$. $A$ generates a random number $x'$ and computes $X^* = g^{x'} \times PW_C'$. Then $A$ impersonates $C$ to sends $M_1 = \{ID_C, X^*\}$ to $G$.

**Step2**. Upon receiving the message $M_1$, $G$ sends $M_2 = \{ID_C, ID_G, X^*\}$ to the server $S$.

**Step 3**: Upon receiving the message $M_2$, the $S$ generates a random number $s$, and computes $\overline{X} = (X^* / H(ID_C, ID_G, pw_C))^s$, $h = g^s$ and $\pi_1 = NIZKPLD(X^*; g, h)$. Then

$S$ sends $M_3 = \{\overline{X}, h, \pi_1\}$ to $G$.

***Step 4***: When $G$ receives $M_3$, he/she generates a random number $y$ and computes

$\overline{Y} = h^y$, $\pi_2 = NIZKPLD(X^*; g, \overline{Y})$, $K_G = (\overline{X})^y$, $AuthG = h_2(ID_C, ID_G, X^*, \overline{Y}, K_G)$

and the session key $sk = h_1(ID_C, ID_G, X^*, \overline{Y}, K_G)$. Then $G$ sends

$M_4 = \{ID_G, h, \overline{Y}, AuthG, \pi_1, \pi_2\}$ to $C$.

***Step 5***: $A$ intercepts the message $M_4$, $C$ computes $K_C = (\overline{Y})^x$ and checks weather $AuthG$ equals $h_2(ID_C, ID_G, X^*, \overline{Y}, K_C)$. If $AuthG$ equals $h_2(ID_C, ID_G, X^*, \overline{Y}, K_C)$, $A$ find the correct password. Otherwise, $A$ repeats step 1), 2), 3), 4) and 5) until find the correct password.

It is clear that if $pw_C'$ equals $pw_C$, then $PW_C' = H(ID_C, ID_G, pw_C)$, $AuthG = h_2(ID_C, ID_G, X^*, \overline{Y}, K_C)$, since

$$\begin{aligned} K_G &= (\overline{X})^y \\ &= ((X^* / H(ID_C, ID_G, pw_C))^s)^y \\ &= (((g^{x'} \times PW_C') / H(ID_C, ID_G, pw_C))^s)^y \\ &= (g^{x'})^{sy} = (g^{sy})^{x'} = (h^y)^{x'} = (\overline{Y})^{x'} = K_C \end{aligned}$$

From the description of the attack we know that Abdalla et al.'s scheme [5] does not prevent the leakage of information of the password from the malicious client $A$. In addition, the attack can be used to attack Abdalla et al.'s another scheme [1].

### 3.2. Session-Key Problem

As in the definitions in [9], a key agreement scheme is said to provide the explicit key confirmation if one entity is assured that the second entity has actually computed the session key. The scheme provides the implicit key confirmation if one entity is assured that the second entity can compute the session key. Note that the property of the implicit key confirmation does not necessarily mean that one entity is assured of the second entity actually possessing the session key. In many applications, it is highly desirable for a key agreement scheme to provide the explicit key confirmation. We can see that the scheme of Abdalla et al. [5] merely provides the implicit key confirmation, because $G$ cannot confirm $C$ has correctly computed the session key after the log-in phase.

## 4. Countermeasure

The vulnerability to the undetectable on-line dictionary attack described above actually stems from an absence of authentication of message in the scheme. To remedy this vulnerability, we can use the method proposed by Byun et al.[2]. First, we let a two party password-based authenticated key exchange (2-PAKE) scheme be executed between $C$ and $S$ in order to generate a session key $sk$. Then we let $C$ create a message authentication code (MAC) of $X^*$ using $sk$. Then, $S$ can check the validity of the $X^*$ through checking MAC of $X^*$ and find the undetectable on-line dictionary attack. However, the execution of the 2-PAKE can increase the burden of the server, the gateway and the client heavily. So, Byun et al.'s method can not be applied in practice. In fact, we just let Abdalla et al.'s scheme provide the implicit key confirmation in order to eliminate the security vulnerability. We modify Abdalla et al.'s [5] scheme as follows.

In our modified scheme, $G$ requires $C$ provide the key confirmation by offering $AuthC$. If malicious client $A$ carry out the undetectable on-line dictionary attack described in section 3.1, $G$ will find the attack, since $A$ can't offer the correct $AuthC$.

**Step 1**: $C$ chooses two random numbers $x$ and $r_C$. Then $C$ computes $X^* = g^x \times H(ID_C, ID_G, pw_C)$, then sends $M_1 = \{ID_C, X^*\}$ to $G$.

**Step 2**: Upon receiving the message $M_1$, $G$ sends $M_2 = \{ID_C, ID_G, X^*\}$ to the server $S$.

**Step 3**: Upon receiving the message $M_2$, the $S$ generates a random number $s$, and computes $\overline{X} = (X^* / H(ID_C, ID_G, pw_C))^s$, $h = g^s$ and $\pi_1 = NIZKPLD(X^*; g, h)$. Then $S$ sends $M_3 = \{\overline{X}, h, \pi_1\}$ to $G$.

**Step 4**: When $G$ receives $M_3$, he/she generates a random number $y$ and computes $\overline{Y} = h^y$, $\pi_2 = NIZKPLD(X^*; g, \overline{Y})$, $K_G = (\overline{X})^y$, and $AuthG = h_2(ID_C, ID_G, X^*, \overline{Y}, K_G)$. Then $G$ sends $M_4 = \{ID_G, h, \overline{Y}, AuthG, \pi_1, \pi_2\}$ to $C$.

**Step 5**: After receiving $M_4$, $C$ computes $K_C = (\overline{Y})^x$ and checks weather $AuthG$ equals $h_2(ID_C, ID_G, X^*, \overline{Y}, K_C)$. If not, $C$ stops the session. Otherwise, $C$ checks weather both of $\pi_1, \pi_2$ is valid. If not, $C$ stops the session, else $C$ computes the session key $sk_C = h_1(ID_C, ID_G, X^*, \overline{Y}, K_C)$ and $AuthC = h_2(ID_G, ID_C, X^*, \overline{Y}, K_C)$. Then $C$ sends the message $M_5 = \{AuthC\}$ to $S$.

**Step 6**: After receiving $M_5$, $S$ checks weather $AuthC$ equals $h_2(ID_G, ID_C, X^*, \overline{Y}, K_G)$. If not $S$ stops the session, else $S$ computes the session key $sk = h_1(ID_C, ID_G, X^*, \overline{Y}, K_G)$.

# 5. Conclusion

Very recently, Abdalla et al. [5] present a new variant of the GPAKE scheme of Abdalla et al. [1]. However, we find that the new scheme is vulnerable to an undetectable on-line guessing attack and can not provide the implicit key confirmation. We also proposed a countermeasure for the security vulnerability.